\documentclass[runningheads]{llncs}
\usepackage[utf8]{inputenc}
\usepackage{tikz}

\usepackage[utf8]{inputenc}
\usepackage{amsmath}
\usepackage{amsfonts}
\usepackage{amssymb}    
\usepackage{dsfont}
\usepackage{mathrsfs}
\usepackage{pxfonts}
\usepackage{pifont}
\usepackage{diagbox}
\usepackage{slashbox,pict2e}
\usepackage{graphicx}
\usepackage{wrapfig}
\usepackage{float}
\usepackage{marvosym}
\usepackage{tikz}
\usepackage{wrapfig}
\usepackage{subfig}
\usepackage{tabularx}
\usepackage[linesnumbered,ruled,vlined]{algorithm2e}
\usepackage{pgfplots}

\newcommand*\circled[1]{\tikz[baseline=(char.base)]{
\node[shape=circle,draw,inner sep=2pt] (char) {#1};}}
\pgfplotsset{compat=1.7}

\title{A Privacy-Preserving Blockchain-based E-voting System}

\author{ %
    Arnab Mukherjee\inst{1}, 
Souvik Majumdar\inst{2}, 
Anup Kumar Kolya\inst{2}
Saborni Nandi\inst{2}\\
} %

\institute{
 Indian Institute of Technology Patna, India\\
 \email{arnab\_2213cs01@iitp.ac.in}\\
\and
RCC Institute of Information Technology, Kolkata, India\\
 \email{\{souvik.majumdar, anupkumar.kolya\}@rcciit.org.in,}
 \email{saborni.nandi@gmail.com}
 } 

\begin{document}

\maketitle

\begin{abstract}
Within a modern democratic nation, elections play a significant role in the nation's functioning. However, with the existing infrastructure for conducting elections using Electronic Voting Systems (EVMs), many loopholes exist, which illegitimate entities might leverage to cast false votes or even tamper with the EVMs after the voting session is complete. The need of the hour is to introduce a robust, auditable, transparent, and tamper-proof e-voting system, enabling a more reliable and fair election process. To address such concerns, we propose a novel solution for blockchain-based e-voting, focusing on the security and privacy aspects of the e-voting process. We consider the security risks and loopholes and aim to preserve the anonymity of the voters while ensuring that illegitimate votes are properly handled. Additionally, we develop a prototype as a proof of concept using the Ethereum blockchain platform. Finally, we perform experiments to demonstrate the performance of the system. 
\end{abstract}

\section{Introduction}
Within a modern democratic nation, elections play a significant role in the functioning of the nation. However, with the existing infrastructure for conducting elections using Electronic Voting Systems (EVMs), many loopholes exist, which illegitimate entities might leverage to cast false votes or even tamper with the EVMs after the voting session is complete. The need of the hour is to introduce a robust, auditable, transparent, and tamper-proof e-voting system, enabling a more reliable and fair election process.


Blockchain \cite{nakamotobitcoin} has become a ground-breaking technology that has transformed how we communicate information over the last few years. Although blockchain is best known for helping the development of digital currencies like Bitcoin, the technology is now being evaluated in a wide range of applications, including insurance, supply chain, health care, e-governance, and many more. According to \cite{survey3}, the main reasons for this success are data immutability, privacy, transparency, decentralization, and distributed ledgers. Most significantly, this technology is strong enough to create trustworthy systems in an unreliable environment thanks to smart contracts' support for business logic and blockchain's ability to execute them. Blockchain technology is ideal for people seeking answers because of the above characteristics and the lack of centralized authority. Most significantly, this technology is strong enough to create trustworthy systems in an unreliable environment since it supports business logic in the form of smart contracts and executes those contracts on the blockchain. Blockchain technology is an ideal solution for systems where several untrusted entities are engaged due to the lack of centralized authority and the characteristics mentioned earlier. 

With the introduction of several proposals for e-voting systems, which include proposals that consider a centralized architecture, as well as distributed systems considering blockchain and distributed databases, a lot of work has been going on in this direction. In \cite{zhaochan}, Zhao et al. proposed a reward or penalty scheme based on the voters’ behavior, marking  it as the first attempt to combine e-voting with blockchain. Lee et al. in \cite{lee} proposed an e-voting protocol involving a Trusted Third Party (TTP) within a blockchain to preserve voters’ choices. Voters are restricted based on their rights within the organization. Here the voters have been identified with a unique identifier, such as Social Security Number (SSN). An authenticated person is traced back to a single candidate. In \cite{end2end}, Bistarelli et al. presented an e-voting protocol based on Bitcoin. Their proposal is based on three phases of voting. Yi Liu et al. in \cite{Liu2017AnEP} proposed a blockchain-based protocol for secure e-voting. To preserve voters’ choices blind signature is used. The blind signature, introduced in \cite{chaum1983blind} by David Chaum, disguises the content of a message before it is signed. Similar to a digital signature, it can be compared with the original unblinded message.

Although these proposals aim at improving the trust, audibility, and transparency of the election process, their proposals fail to preserve the private and identifiable information of the voters. In this paper, we propose a novel solution for blockchain-based e-voting, focusing on the security and privacy aspects of the e-voting process. We consider the security risks and loopholes and aim to preserve the anonymity of the voters while ensuring that illegitimate votes are properly handled. 

To the best of our knowledge, this proposal is one of its kind, leveraging blockchain technology and zero-knowledge proof to ensure a secure and reliable e-voting platform.

To summarize, the main contributions of this paper are:
\begin{itemize}
    \item We present a comprehensive method for trustworthy and secure electronic voting that makes use of blockchain technology, zero-knowledge proofs, and smart contracts. By taking into account all stakeholders and the beneficial services that are pertinent to them, we assess the worst-case scenario. 
    \item By implementing suitable access control and authentication systems, we address security and privacy concerns. 
    \item As proof of concept, we create a prototype on the Ethereum blockchain. 
    \item Finally, we perform experiments to demonstrate the performance of the system. 
\end{itemize}

\noindent The structure of the paper is organized as follows: In section \ref{sec-relworks}, we discuss some of the state-of-the-art proposals and perform a comparative analysis among them. With section \ref{sec-prelims}, we briefly recall the basics of the technologies and concepts we leverage in our proposal. Section \ref{sec-approach} describes our proposed blockchain-based e-voting system. In Section \ref{sec-proof}, we present the proof of concept using the Ethereum platform. The experimental results are discussed in Section \ref{sec-expres}. Finally, we conclude our work with Section \ref{sec-con}.

\begin{figure*}[ht]
\centering
\includegraphics[width=0.95\textwidth]{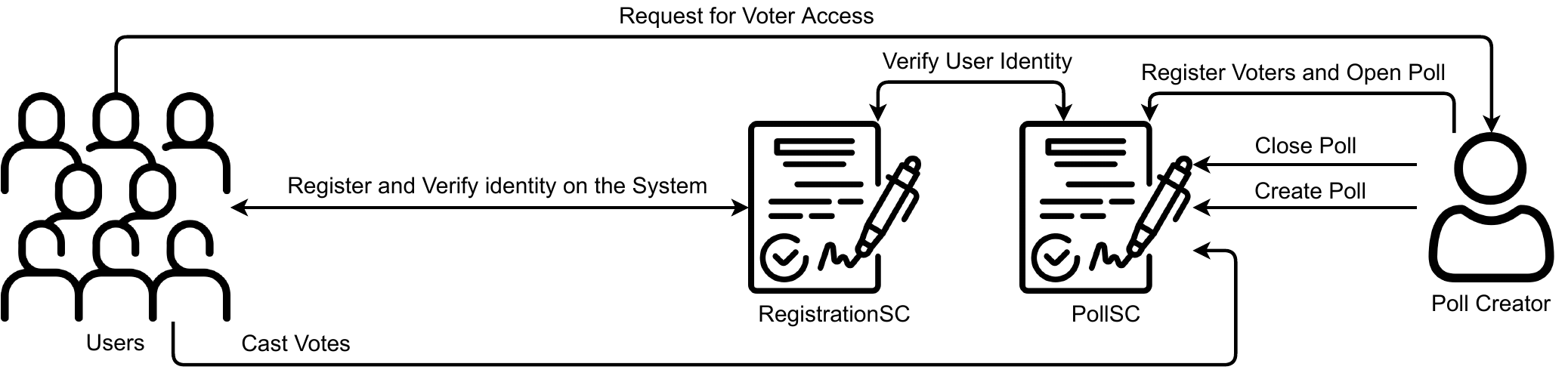}
\caption{Overall architecture of the proposed system }
\label{fig:proposal}
\end{figure*}

\section{Related Works}\label{sec-relworks}
Let us now discuss some of the state-of-the-art proposals for e-voting. Our literature survey revealed some of the major proposals for e-voting systems, including centralized and blockchain-based proposals. 

In \cite{zhaochan}, Zhao et al. proposed a reward or penalty scheme based on the voters’ behavior, marking  it as the first attempt to combine e-voting with blockchain. Their work was based on Bitcoin and satisfies the basic features of a blockchain-based proposal, i.e., privacy, verifiability, and auditability. Similar to \cite{zhaochan}, in 2016, Lee et al. in \cite{lee} proposed an e-voting protocol involving a TTP (trusted third party) within a blockchain network for preserving voters’ choices. Additionally, voters are restricted based on their rights within their own organizations. This paper identifies voters with identification numbers such as SSN (Social security number), leading an authenticated person to be tracked back to a single candidate. 

Bistarelli et al. in \cite{end2end} proposed an e-voting protocol using the Bitcoin platform. As per their proposal, the e-voting process consists of three phases, i.e., the Pre-voting phase, including the candidate nomination and registration process, and the voter registration phase. Next is the voting phase, where the voters cast their votes. Finally, when the voting process is terminated, and the results of the election are published. The complete system is based on the Bitcoin platform, and the stakeholders are identified using their public keys/wallet addresses.

Liu et al. in \cite{Liu2017AnEP} proposed an e-voting protocol without the involvement of a third party. The protocol has been designed based on blind signatures and blockchain. To preserve the anonymity of voters’ choices, a blind signature is used. The blind signature was initially introduced by David Chaum. The content of a message is disguised before it is signed. In \cite{chaum1983blind}, Yung et al. proposed an e-voting system without the involvement of a trusted third party. However, their proposal requires extensive computation to verify and commit transactions, making the proposal very inefficient and slow.

Much research has been going on in this domain of e-voting systems, but the existing proposal doesn't address the privacy and security concerns involving an e-voting system. The blockchain-based e-voting systems aim at bridging the gaps of traditional centralized e-voting systems. However, they don't address issues like non-repudiation, privacy, and anonymity of voters, concerning authentication and verification of votes. To this aim, we propose a novel blockchain-based e-voting system, with the aid of zero-knowledge proofs to mitigate the above-mentioned shortcomings.

\section{Preliminaries}\label{sec-prelims}
Let us now briefly recall from \cite{nakamotobitcoin,smt,IPFS,SW,FC,zk1,zk3}, the preliminary concepts of Blockchain Technology, Decentralized File System, and Zero-Knowledge Protocol, which we will often refer to throughout the paper.

\paragraph*{Blockchain and Smart Contracts \cite{nakamotobitcoin,smt}.}
Blockchain technology \cite{nakamotobitcoin}, in the last decade, has emerged as one of the most cutting-edge technologies, creating a solid platform for safe, open, and immutable information sharing. For maintaining a permanent and tamper-proof record of transactions made on the decentralized peer-to-peer network, blockchain is a distributed, transparent, trustless, and publicly available ledger. In recent years, both businesses and academics have paid intense attention to these unique characteristics of blockchain technology. 

A new addition to blockchain technology is the addition of support for smart contracts \cite{smt}. High-level Turing complete languages like Solidity, Rust, or Go are used to create smart contracts, which are functions that may be executed on the blockchain. By automatically carrying out and enforcing the terms of an agreement, smart contracts help eliminate any middlemen and unreliable third parties between the participating blockchain network participants. 

\paragraph*{Decentralized File System \cite{IPFS,FC,SW}.}
Decentralized storage is a peer-to-peer file-sharing network with a platform for shared global memory that offers the system a potential remedy for resilience and efficiency. Thanks to decentralized storage, files may be stored and retrieved from a decentralized location. In recent years, several decentralized file systems have been introduced, such as IPFS \cite{IPFS}, Filecoin \cite{FC}, and Storj \cite{SW}, with several attractive features, including greater security and privacy, low-cost of maintenance, verifiability, transparency, content addressing, and high flexibility. 

\paragraph*{Zero Knowledge Proof Protocol \cite{zk1,zk3}.}
Zero-Knowledge Proof is an encryption and authentication system that enables one party (Prover) to demonstrate to another party (Verifier) that a Common Reference String (CRS) is genuine and legitimate without disclosing any private information other than that required to validate the statement itself. The main characteristics that make this protocol popular are completeness, soundness, and zero knowledge. Zero-knowledge proofs usually come in two different flavors. There are two types of zero-knowledge proofs: the Interactive Zero-Knowledge Proof, where the Prover uses a series of actions based on mathematical probability principles to persuade the Verifier of a specific fact, and the Non-Interactive Zero-Knowledge Proof, where the Prover generates all of the challenges at once and the Verifier(s) can respond later, limiting the possibility of collusion.

Our proposal proposes an identification system that employs Succinct Non-Interactive Zero-Knowledge Proofs (zk-SNARKs) to demonstrate the identity without disclosing any personal or sensitive information about the stakeholders (Prover). By combinations of elliptic curves and homomorphic encryption, zk-SNARKs provide genuine proofs without knowing the evaluated point. 

\section{Proposed Approach}\label{sec-approach}

In this section, we elucidate our proposal for the blockchain-based e-voting system. Since our primary objective is to maintain privacy while ensuring the security and authenticity of votes throughout the entire election process (commencing from election registration to revealing results), we consider the system architecture depicted in Figure \ref{fig:proposal}. The proposed model consists of the following stakeholders: \circled{1}, the poll creator, and \circled{2}, the voters. The process starts with the poll creator initiating a poll and registering the addresses of the voters to the poll using the e-voting smart contract. Next, the voters are made aware of the poll, and the process proceeds with them casting their votes. Once all of the registered voters have cast their votes, the poll is automatically closed, and the e-voting smart contract reveals the poll results. Since all of the transactions performed, including the creation of the poll, registering of the voters, the casting of votes, and the publication of poll results, are facilitated through the e-voting smart contract on the blockchain network, every transaction performed is recorded on the distributed ledger, ensuring reliable and auditable access to the poll, and enforcing trust within the system involving untrusted parties. The recap, the overall processes of the system involve the following phases:

\begin{enumerate}
    \item Registration of stakeholders, i.e., Poll Creator and the Voters.
    \item Creation of the Poll event.
    \item Registration of the respective voters to the Poll.
    \item Voting process involves the voters casting their votes using the e-voting smart contract.
    \item Closing of the Poll.
    \item Publication of the Poll results.
\end{enumerate}

\noindent Let us not describe each of these phases in detail.

\subsection{Registration \& Verification of Stakeholders}

This phase allows a legitimate person to register as a stakeholder and perform activities in the system. We propose a registration and verification mechanism leveraging zero-knowledge proofs (ZKP) to ensure the same. The process starts with legitimate users $U_i$ approaching the system and executing the following actions in sequence:

\begin{itemize}
    \item The user $U_i$ sends a registration Request $Rq_i$ to the Identity Verifier $V_j$, to register on the platform. This action generates a temporary key $TmpKey_i$, for $U_i$. Here $TmpKey_i$ is generated randomly based on device credentials and timestamp. $TmpKey_i$ and $Rq_i$ are sent over a secure HTTPS channel to $V_j$.

    \item Extraction of $TmpKey_i$ and manifesting of $Rq_i$ occur once $V_j$ has received $Rq_i$. Then, $V_j$ creates an unverified identification called $UnvID_i$. This is given to user $U_i$ and is only confirmed after completing the ZKP procedure.     

    \item For the next step, a signal flag $Sig_{ZKP}$ is forwarded to $U_i$, indicating the start of the ZKP process.

    \item The prover and verifier are initialized as $U_i$ and $V_j$, respectively, in the ZKP process. Without disclosing sensitive information about the prover to the verifier, this approach aids in confirming the veracity of the prover.     

    \item Next, $U_i$ determines the value of $r$, or a random number, which will later be utilized to determine the value of $d$. This $d$ is now transmitted across a secure channel to the verifier $V_j$. 

    \item The $V_j$ extracts $d$ and selects two challenges for the prover. These challenges are sent to the prover $U_i$ over a secure channel.

    \item On receiving the challenges, $U_i$ computes the answers for them and transmits the answers $A_i$ over a secure channel to $V_j$.

    \item Finally, after $V_j$ receives the answers, they verify them, and once the verification is complete, a permanent identity $ID_i$ is generated for user $U_i$.
\end{itemize}

\subsection{Creation of the Poll Event}
With this phase of the system, a registered user $U_x$ is allowed to create a poll. This process starts with the invocation of the \texttt{createPoll()} method of \textsf{PollSC} while providing the necessary details. These details include a poll name $PollName_x$, a description of the poll $PollDesc_x$, and a list of poll options or choices denoted by $PollOpts_{<x,i>}$. After this is successfully done, a poll ID is returned as $Poll_x$. Next, the user $U_x$ is required to register the respective users $U_i$ to the poll so that $U_i$ can cast their votes on the available options $PollOpts_{x,i}$. 

\subsection{Registration of Voters to the $Poll_x$}
In this phase, the poll creator $U_x$ invokes the \texttt{registerVoters()}, providing an array of registered users $[U_1, U_2, .... U_j]$ so that they are added to the poll $Poll_x$. Without this step, no user registered on the system $U_i$ can cast their votes. With the successful completion of this phase, during a user $U_i$ casting their vote, the \textsf{PollSC} will be verified if they are registered to the respective poll event $Poll_x$. Only their votes will be considered valid if they are registered to that respective poll $Poll_x$. However, if the poll creator wants their poll to be an open poll, then they need to invoke the \texttt{setOpen()} method instead to make the $Poll_x$ open to all registered users on the system. Next, the user $U_x$ is required to invoke the \texttt{openPoll()} method, which will allow the $Poll_x$ to accept votes from registered and verified users $U_i$, which is manifested in the subsequent sections. 

\subsection{Casting of Votes}
Once the poll creator $U_x$ opens up the poll invoking the \texttt{openPoll()} method, all registered users $[...U_j]$ on that respective poll $Poll_x$ will be able to cast their votes for the available poll options $<PollOpt_1, PollOpt_2,...PollOpt_y>$. The registered voter $U_i$ need to invoke the \texttt{castVotes()} supplying the required parameters $<PollID, PollChoiceIndex>$. Once the casting of votes is successful, granted that the required checks were successful, i.e., voter $U_i$ was registered to the $Poll_x$, and that $U_i$'s identity was verified in the Registration and Verification phase, their vote $Vote_m$ will be registered on the \textsf{PollSC} smart contract, for $Poll_x$. And this vote will be appended to the blockchain ledger permanently. 

\subsection{Closing of the Poll}
After all the registered voters cast their votes on $Poll_x$, the poll creator can close the poll, invoking the \texttt{closePoll()} method of the \textsf{PollSC} smart contract. Doing so will render the poll closed, i.e., no more votes can be cast, and the vote results will be declared after computing them and storing them in the smart contract. Closing the poll involves setting the $pollStatus$ to $false$, which, when checked against before accepting votes, will render the \texttt{castVotes()} method to fail.

\subsection{Publication of Poll Results}
Once the poll is closed by invoking the \texttt{closePoll()} method, the poll results will be available to all the participants and stakeholders in the system. To access the poll results, the \texttt{getPollResults()} method needs to be invoked from the \textsf{PollSC} smart contract, providing the $PollID$ as an argument to the method. This method invocation will return the winning choice $PollOpt_y$ if the poll is closed, else will return \texttt{null}.

\section{Proof of Concept}\label{sec-proof}
This section summarises the experimental findings and describes the prototype implementation of the proposed system. The Ethereum blockchain platform serves as the foundation for the system. The following elements make up the prototype implementation: 

\begin{enumerate}
    \item The system under consideration is built on the Ethereum blockchain network. Our approach for deploying smart contracts makes use of the Rinkeby Ethereum testnet \footnote{https://www.rinkeby.io}. 
    \item To enable various system functionalities, a collection of smart contracts written in Solidity are deployed. 
    \item Here, we employ an instance of the InterPlanetary File System (IPFS)\footnote{https://ipfs.io/}, a peer-to-peer storage system to store documents like registration and verification reports. 
    \item We create Solidity smart contracts to create and validate zero-knowledge proofs, using the ZoKrates toolbox\footnote{https://zokrates.github.io} for the zkSNARKs implementation on Ethereum. 
\end{enumerate}

Now, let's quickly go through each of the two smart contracts' functionalities that were implemented in our prototype: 

\begin{enumerate}
    \item \textbf{RegistrationSC:} This smart contract handles the whole process of registering stakeholders in the system.  
    \item \textbf{PollSC:} This smart contract is responsible for all activities related to the creation of polls, voter registration, voting, closing of polls, and declaration of poll results.
\end{enumerate}

By using the ZoKrate plugin in the Remix IDE \footnote{https://remix.ethereum.org/} and creating the circuit in \texttt{.zok} files, we implement zk-SNARK in our system. The zk-SNARK prototype uses a smart contract and an off-chain proof generator to validate the proof. 

\subsection{Experimental Results}\label{sec-expres}
In this section, let us manifest the experimental evaluations we conducted to verify the practicality of our proposal. We conduct the experiments on a machine equipped with a 24-core Intel Xeon CPU running at a base frequency of 1.99 GHz, 128 GB of RAM, and the Ubuntu 18.04 LTS 64-bit operating system. 

In the first round of experiments, we keep track of transaction costs, i.e., the gas costs of the various system operations.
Gas costs are divided into two categories: (1) Gas costs associated with deploying a smart contract and (2) Gas costs associated with executing various smart contract operations.

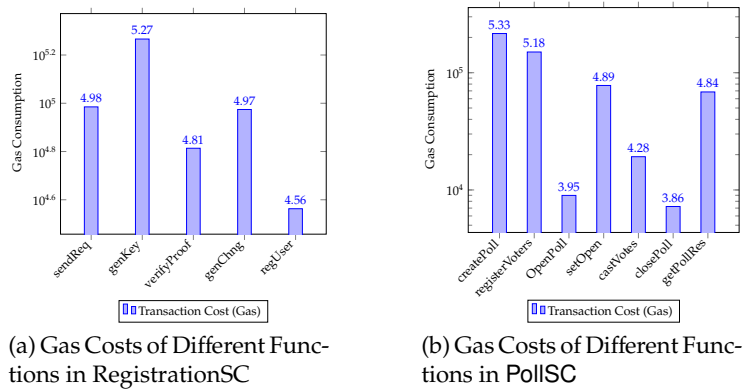
\begin{figure}[ht]
\centering
\subfloat[Gas Costs of Different Functions in RegistrationSC]{\label{regt}%
  \resizebox{0.35\textwidth}{!}{%
  \begin{tikzpicture}
\begin{axis}[
   ybar=6pt,
    enlargelimits=0.15,
    legend style={at={(0.5,-0.15)},
      anchor=north,legend columns=-1},
    ylabel={\ Gas Consumption},
    legend style={at={(0.5,-0.3)},anchor=north,legend cell align=left},
    x tick label style={rotate=45, anchor=east},
    symbolic x coords={sendReq,genKey,verifyProof,genChng,regUser},
    xtick=data,
    nodes near coords,
    nodes near coords align={vertical},
    ymode=log,
    log basis y={10},
    ]
\addplot coordinates {(sendReq,96584) (genKey,184584)(verifyProof,65082)(genChng,94214)(regUser,36548)};
\legend{Transaction Cost (Gas)}
\end{axis}
\end{tikzpicture}
}%
}\hfil
\subfloat[Gas Costs of Different Functions in \textsf{PollSC}]{\label{tokt}%
  \resizebox{0.35\textwidth}{!}{%
\begin{tikzpicture}
\begin{axis}[
   ybar=6pt,
    enlargelimits=0.15,
    legend style={at={(0.5,-0.15)},
      anchor=north,legend columns=-1},
    ylabel={\ Gas Consumption},
    legend style={at={(0.5,-0.3)},anchor=north,legend cell align=left},
    x tick label style={rotate=45, anchor=east},
    symbolic x coords={createPoll,registerVoters,OpenPoll,setOpen,castVotes,closePoll,getPollRes},
    xtick=data,
    nodes near coords,
    nodes near coords align={vertical},
    ymode=log,
    log basis y={10},
    ]
\addplot coordinates {(createPoll,215949) (registerVoters,150240)(OpenPoll,8981)(setOpen,77852)(castVotes,19214)(closePoll,7223)(getPollRes,68547)};
\legend{Transaction Cost (Gas)}
\end{axis}
\end{tikzpicture}
}%

}
\caption{Gas Costs of Different Functions of Different Smart Contracts }
\label{fig_bench}
\vspace{-0.1in}
\end{figure}

Figure \ref{fig_bench} represents the Gas cost with respect to different function invocations in RegistrationSC and PollSC smart contracts.

Let's now talk about the experimental findings from the second stage of our study, which was performed to test the implementation of our prototype.
We assess the stability and functionality of our system, specifically under load. Utilizing the Hyperledger Caliper benchmarking tool this is accomplished.

\begin{figure}[ht]
    \centering
    \scriptsize 
    \subfloat[Read Operation]{
        \label{caliper_read}\resizebox{0.35\textwidth}{!}{
        \begin{tikzpicture}
        	\begin{axis}  
        		[  
        			ybar, 
        			axis y line*=left,
        			enlargelimits=0.15,  
        			ylabel={\ref{plot1} Throughput (TPS)},
        			xlabel={Send Rate (TPS)},  
        			symbolic x coords={250,500,750,1000,1250,1500,1750,2000,2250,2500,2750,3000},  
        			xtick=data,
        			nodes near coords align={vertical},
        			xticklabel style={rotate=90},
        		]  
            		\addplot coordinates {(250,250) (500,500) (750,750) (1000,999.7) (1250,1240.6) (1500,1420.7) (1750,1732.1) (2000,1999.6) (2250,2243.7) (2500,2499.7) (2750,2741.5) (3000,2984.2)};  
        	\end{axis}
        	\begin{axis}  
        		[  
        		    axis x line=none,
        		    axis y line*=right,
        			enlargelimits=0.15,
        			ylabel={\ref{plot2} Average Latency (s)},
        		]  
        		\addplot[color=red,mark=square] coordinates {(250,0.01) (500,0.01) (750,0.01) (1000,0.01) (1250,0.01) (1500,0.01) (1750,0.01) (2000,0.01) (2250,0.01) (2500,0.01) (2750,0.01) (3000,0.01)};
        	\end{axis}
        \end{tikzpicture}
        }
    }
    \hfill
    \subfloat[Write Operation]{
        \label{caliper_write}\resizebox{0.35\textwidth}{!}{
        \begin{tikzpicture}
        	\begin{axis}  
        		[  
        			ybar,  
        			axis y line*=left,
        			enlargelimits=0.15,  
        			ylabel={\ref{plot1} Throughput (TPS)},
        			xlabel={Send Rate (TPS)},  
        			symbolic x coords={25,50,75,100,125,150,175,200,225,250,275,300},  
        			xtick=data,
        			nodes near coords align={vertical},
        			xticklabel style={rotate=90},
        		]
        		\addplot coordinates {(25,19.0) (50,21.5) (75,22.7) (100,24.7) (125,25.9) (150,27.0) (175,28.8) (200,30.4) (225,31.2) (250,34.0) (275,36.2) (300,37.1)};
                \label{plot1}
        	\end{axis}
        	\begin{axis}  
        		[  
        		    axis x line=none,
        		    axis y line*=right,
        			enlargelimits=0.15,
        			ylabel={\ref{plot2} Average Latency (s)}, 
        		]  
            		\addplot[color=red,mark=square] coordinates {(25,5.7) (50,15.73) (75,24.61) (100,33.42) (125,41.16) (150,48.8) (175,52.38) (200,57.89) (225,62.15) (250,65.68) (275,67.8) (300,72.54)};
                \label{plot2}
        	\end{axis}
        \end{tikzpicture} 
        }
    }
    \caption{Caliper Ethereum Benchmark Plots}
    \label{fig_caliper}
\end{figure}
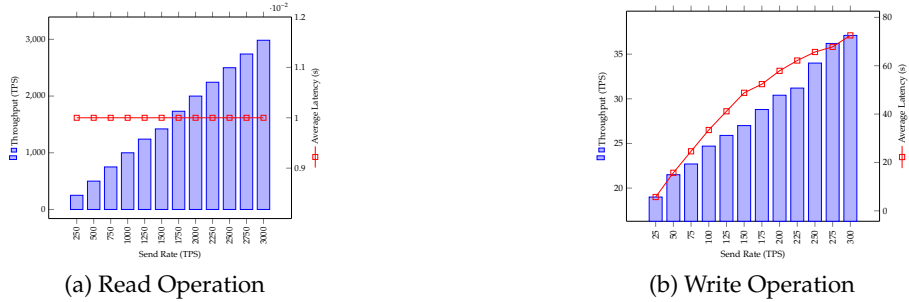

To accomplish this, we performed Hyperledger Caliper benchmarks for the metrics \texttt{send rate} (total transaction requests sent per second) and \texttt{concurrent users} (number of concurrent users at an instance). 
With intervals of 25 TPS, we changed the \texttt{transmission rate} from 25 Transactions Per Second (TPS) to 300 TPS. We altered the number of users for \texttt{concurrent users} from 25 to 100 with a 25-user interval. We recorded the system's average latency and throughput for a particular ratio of \texttt{concurrent users} and \texttt{transmit rate} for evaluation metrics. Figure \ref{fig_caliper} shows the typical latency and transaction throughput.

\section{Conclusion}\label{sec-con}
In the paper, employing the benefits of blockchain technology and smart contracts, we suggest an innovative approach to managing e-waste.
We analyze the possible scopes of improvement in the overall e-voting system, considering the honesty of users, traceability of actions, and privacy. We provide privacy-focused solutions for the system's information sharing and validation by using zero-knowledge-proof protocols. In addition, we develop and apply the necessary access control techniques to prevent unwanted access to critical data within the system. The positive experimental findings show smart contracts perform efficiently and affordably throughout various phases.

For future scopes of work, we will consider integrating AI into the system, making it easier to do different vote management duties to make enhancements.

\bibliographystyle{IEEEtran}
\bibliography{thebibliography}

\begin{thebibliography}{10}
\providecommand{\url}[1]{#1}
\csname url@samestyle\endcsname
\providecommand{\newblock}{\relax}
\providecommand{\bibinfo}[2]{#2}
\providecommand{\BIBentrySTDinterwordspacing}{\spaceskip=0pt\relax}
\providecommand{\BIBentryALTinterwordstretchfactor}{4}
\providecommand{\BIBentryALTinterwordspacing}{\spaceskip=\fontdimen2\font plus
\BIBentryALTinterwordstretchfactor\fontdimen3\font minus
  \fontdimen4\font\relax}
\providecommand{\BIBforeignlanguage}[2]{{%
\expandafter\ifx\csname l@#1\endcsname\relax
\typeout{** WARNING: IEEEtran.bst: No hyphenation pattern has been}%
\typeout{** loaded for the language `#1'. Using the pattern for}%
\typeout{** the default language instead.}%
\else
\language=\csname l@#1\endcsname
\fi
#2}}
\providecommand{\BIBdecl}{\relax}
\BIBdecl

\bibitem{nakamotobitcoin}
S.~Nakamoto, ``Bitcoin: A peer-to-peer electronic cash system,'' in \emph{.},
  2008.

\bibitem{survey3}
Z.~Zheng, S.~Xie, H.~Dai, X.~Chen, and H.~Wang, ``Blockchain challenges and
  opportunities: a survey,'' \emph{International Journal of Web and Grid
  Services}, vol.~14, pp. 352--375, 2018.

\bibitem{zhaochan}
Z.~Zhao and T.-H.~H. Chan, ``How to vote privately using bitcoin,'' in
  \emph{Information and Communications Security}.\hskip 1em plus 0.5em minus
  0.4em\relax Springer International Publishing, 2016, pp. 82--96.

\bibitem{lee}
L.~K. J. J. I. E.~T. G. and K.~H. J., ``Electronic voting service using
  block-chain,'' \emph{Journal of Digital Forensics, Security and Law},
  vol.~11, 2016.

\bibitem{end2end}
F.~Santini, S.~Bistarelli, I.~Mercanti, and P.~Santancini, ``End-to-end voting
  with non-permissioned and permissioned ledgers,'' \emph{Journal of Grid
  Computing}, vol.~17, 03 2019.

\bibitem{Liu2017AnEP}
Y.~Liu and Q.~Wang, ``An e-voting protocol based on blockchain,'' \emph{IACR
  Cryptol. ePrint Arch.}, vol. 2017, p. 1043, 2017.

\bibitem{chaum1983blind}
D.~Chaum, ``Blind signatures for untraceable payments,'' in \emph{Advances in
  cryptology}.\hskip 1em plus 0.5em minus 0.4em\relax Springer, 1983, pp.
  199--203.

\bibitem{smt}
J.~{Liu} and Z.~{Liu}, ``A survey on security verification of blockchain smart
  contracts,'' \emph{IEEE Access}, pp. 77\,894--77\,904, 2019.

\bibitem{IPFS}
J.~Benet, ``{IPFS} - content addressed, versioned, {P2P} file system,'' July
  2014, online.

\bibitem{SW}
``{SWARM},'' Available: https://swarm.ethereum.org/.

\bibitem{FC}
``{FILECOIN},'' Available: https://filecoin.io/.

\bibitem{zk1}
\BIBentryALTinterwordspacing
M.~Blum, P.~Feldman, and S.~Micali, ``Non-interactive zero-knowledge and its
  applications,'' in \emph{Proceedings of the Twentieth Annual ACM Symposium on
  Theory of Computing}, ser. STOC '88.\hskip 1em plus 0.5em minus 0.4em\relax
  New York, NY, USA: Association for Computing Machinery, 1988, p. 103–112.
  [Online]. Available: \url{https://doi.org/10.1145/62212.62222}
\BIBentrySTDinterwordspacing

\bibitem{zk3}
S.~Bowe, A.~Gabizon, and M.~D. Green, ``A multi-party protocol for constructing
  the public parameters of the pinocchio zk-snark,'' in \emph{Financial
  Cryptography and Data Security}, A.~Zohar, I.~Eyal, V.~Teague, J.~Clark,
  A.~Bracciali, F.~Pintore, and M.~Sala, Eds.\hskip 1em plus 0.5em minus
  0.4em\relax Berlin, Heidelberg: Springer Berlin Heidelberg, 2019, pp. 64--77.

\end{thebibliography}

\end{document}